\documentclass[a4paper,fleqn,usenatbib]{mnras}

\usepackage{newtxtext,newtxmath}

\usepackage[T1]{fontenc}
\usepackage{ae,aecompl}

\usepackage{graphicx}	
\usepackage{amsmath}	
\usepackage{amssymb}	
\usepackage{siunitx}    
\usepackage{bbm} 		
\usepackage[dvipsnames]{xcolor}		
\usepackage{soul} 		
\usepackage{booktabs}	
\usepackage{array}		
\usepackage[inline]{enumitem}	
\usepackage{multirow}   
\usepackage{etoolbox}   
\sethlcolor{yellow}
\usepackage{ragged2e}

\newcommand{\td}{\textrm{d}}
\newcommand{\pd}[2]{\frac{\partial #1}{\partial #2}}
\newcommand{\xc}{X_\textrm{c}}
\newcommand{\avet}{\langle \Delta t \rangle}

\makeatletter
\patchcmd{\NAT@citex}
  {\@citea\NAT@hyper@{%
     \NAT@nmfmt{\NAT@nm}%
     \hyper@natlinkbreak{\NAT@aysep\NAT@spacechar}{\@citeb\@extra@b@citeb}%
     \NAT@date}}
  {\@citea\NAT@nmfmt{\NAT@nm}%
   \NAT@aysep\NAT@spacechar\NAT@hyper@{\NAT@date}}{}{}
\patchcmd{\NAT@citex}
  {\@citea\NAT@hyper@{%
     \NAT@nmfmt{\NAT@nm}%
     \hyper@natlinkbreak{\NAT@spacechar\NAT@@open\if*#1*\else#1\NAT@spacechar\fi}%
       {\@citeb\@extra@b@citeb}%
     \NAT@date}}
  {\@citea\NAT@nmfmt{\NAT@nm}%
   \NAT@spacechar\NAT@@open\if*#1*\else#1\NAT@spacechar\fi\NAT@hyper@{\NAT@date}}
  {}{}
\makeatother

\title[Glitches triggered by a Brownian process]{Long-term statistics of pulsar glitches triggered by a Brownian stress accumulation process}

\author[Carlin et al.]{
J. B. Carlin,$^{1,\,2}$
A. Melatos$^{1,\,2}$
\\
$^{1}$School of Physics, University of Melbourne, Parkville, VIC 3010, Australia\\
$^{2}$Australian Research Council Centre of Excellence for Gravitational Wave Discovery (OzGrav), University of Melbourne, \\ \ Parkville, VIC 3010, Australia\\
}

\date{Accepted XXX. Received YYY; in original form ZZZ}

\pubyear{2019}

\begin{document}
\label{firstpage}
\pagerange{\pageref{firstpage}--\pageref{lastpage}}
\maketitle

\begin{abstract}
A microphysics-agnostic meta-model of rotational glitches in rotation-powered pulsars is developed, wherein the globally averaged internal stress accumulates as a Brownian process between glitches, and a glitch is triggered once a critical threshold is surmounted. Precise, falsifiable predictions are made regarding long-term event statistics in individual pulsars. For example, the Spearman cross-correlation coefficient between the size of a glitch and the waiting time until the next glitch should exceed 0.25 in all pulsars. Among the six pulsars with the most recorded glitches, PSR J0537$-$6910 and PSR J0835$-$4510 are consistent with the predictions of the meta-model, while PSR J1740$-$3015 and PSR J0631$+$1036 are not. PSR J0534$+$2200 and PSR J1341$-$6220 are only consistent with the meta-model, if there exists an undetected population of small glitches with small waiting times, which we do not resolve. The results are compared with a state-dependent Poisson process, another microphysics-agnostic meta-model in the literature. The results are also applied briefly to recent pulse-to-pulse observations of PSRJ0835$-$4510, which appear to reveal evidence for a negative fluctuation in rotation frequency just prior to the 2016 glitch.
\end{abstract}




\section{Introduction}
The secular braking of rotation-powered pulsars is perturbed by two phenomena: glitches and timing noise. Timing noise, or stochastic wandering of the spin frequency, shows up in timing residuals as a red-noise process with an auto-correlation time-scale of days to weeks \citep{Cordes1980, Price2012, Parthasarathy2019}. Glitches are impulsive spin-up events that recur erratically \citep{Melatos2008, Espinoza2011, Fuentes2019}. 

The microphysical mechanism that triggers glitches is an open question. Candidates include superfluid vortex avalanches \citep{Anderson1975, Warszawski2011}, starquakes \citep{Larson2002, Middleditch2006}, hydrodynamic instabilities \citep{Andersson2003, Mastrano2005, Glampedakis2009} and more; see \citet{Haskell2015} for a modern review. Most of these mechanisms are predicated on the idea that the electromagnetic braking of the crust increases stress (e.g. elastic strain or differential rotation) in the system, some fraction of which is released spasmodically at each glitch. If the stress increases deterministically between glitches, the long-term glitch activity can be described by a state-dependent Poisson (SDP) process which links the instantaneous glitch rate to the stress in the system; glitches become more likely as the stress approaches a threshold \citep{Fulgenzi2017}. The SDP process is a meta-model in the sense that it encompasses phenomenologically the stress-release idea at the core of the mechanisms listed above without specializing to the microphysics of the mechanism. It makes falsifiable statistical predictions about long-term observations of the sizes and waiting times of glitches and their correlations \citep{Fulgenzi2017, Melatos2018, Carlin2019, Carlin2019b}. 

Pulse-to-pulse observations of a glitch in the Vela pulsar (PSR J0835$-$4510) were made at the Mount Pleasant radio telescope in December 2016 \citep{Palfreyman2018}. Bayesian analysis finds evidence for a rotational slowdown (``precursor'') immediately prior to the glitch \citep{Ashton2019}. The slowdown is of the same order as the pulse jitter, i.e. pulse-to-pulse variations in the pulse profile, possibly caused by magnetospheric fluctuations unrelated to the internal stress. Another possibility --- certainly not unique --- is that the slowdown represents a random internal (e.g. hydrodynamic) fluctuation, which drives the stress above a critical threshold, triggering the glitch \citep{Ashton2019}. Stochastic fluctuations in the internal stress may be caused by superfluid turbulence, for example \citep{Melatos2007, Melatos2014, Khomenko2019a}. 

In this paper we do not seek to adjudicate on the putative link between internal stochastic fluctuations and an observed rotational slowdown prior to a glitch. Nor do we seek to model such a link directly. Instead, motivated partly by the Vela data, we investigate an alternative to the SDP meta-model, wherein glitches are the result of an internal, unobservable, globally averaged stress that evolves stochastically as a Brownian process, until a glitch is triggered at a critical stress threshold. The Brownian meta-model differs from the SDP meta-model by allowing the stress to evolve stochastically between glitches (instead of increasing deterministically), and triggering a glitch only when a critical threshold is reached (instead of at any time before the threshold is reached). Together the two meta-models encompass a large set of plausible microphysical mechanisms. Both models make falsifiable predictions about long-term statistics, a valuable feature. We describe the details of the Brownian meta-model in Section \ref{sec:bm_meta}. In Section \ref{sec:results} we explore its long-term statistical predictions. In Section \ref{sec:comp} we compare data from the six pulsars with the highest number of recorded glitches with the predictions of the Brownian meta-model, with an eye towards falsification. An analogous study of the SDP meta-model can be found elsewhere \citep{Melatos2018, Carlin2019, Carlin2019b}. In Section \ref{sec:poptrends} we discuss how population trends may inform meta-model parameters.

\section{Brownian stress accumulation}
\label{sec:bm_meta}
\subsection{Equation of motion}
\label{sec:eom}
We define $X$ to be a stochastic variable equal to the globally averaged stress in the system. In the superfluid vortex avalanche picture $X$ is proportional to the lag between the angular speed of the rigid crust and the superfluid interior. In the crustquake picture $X$ is proportional to the elastic strain in the crust.

Between glitches we propose that $X(t)$ evolves according to a Wiener process, which obeys the Langevin (It\^{o}) equation
\begin{equation}
\label{eq:sde}
\frac{\td X(t)}{\td t} = \xi + \sigma B(t)\ \ ,
\end{equation}
with drift coefficient $\xi$ (units: stress$/$time) and diffusion coefficient $\sigma$ $[$units: stress$/({\rm time})^{1/2}]$, and where $B(t)$ is a white noise process of zero mean and unit variance \citep{Cox1965, Gardiner2009}. We assume both $\xi$ and $\sigma$ are constant with time. Practically, at each time step, the stress increments by $\xi$ and undergoes a random step (up or down) by $\sigma$ multiplied by a random number drawn from a Gaussian with zero mean and variance equal to the time step. Equation \eqref{eq:sde} leads to the Fokker-Planck equation 
\begin{equation}
\label{eq:cke}
\pd{p}{t} = -\xi \pd{p}{X} + \frac{\sigma^2}{2} \pd{^2p}{X^2}\ \ ,
\end{equation}
where $p\,\td X = p(X, t \,|\, X_0)\, \td X$ is the probability of finding the stress in the region $(X, X + \td X)$ at time $t$, given that it started at $X=X_0$ after a glitch at $t=0$, viz.
\begin{equation}
\label{eq:ic}
p(X, t=0 \,|\, X_0) = \delta(X - X_0)\ \ .
\end{equation}

The Brownian process terminates at $X=\xc$, i.e. $\xc$ is the stress threshold where a glitch is triggered. The glitch decrements the stress by a random amount $\Delta X$, drawn from a stress-release distribution, discussed in Section \ref{sec:dist}. Mathematically, the termination of the Brownian process at $X=\xc$ corresponds to an absorbing boundary condition:
\begin{equation}
\label{eq:abs}
0 = p(X=\xc, t \,|\, X_0)\ \ .
\end{equation}
We also require $X(t) \geq 0$; the stress is never negative\footnote{In the vortex unpinning picture, for example, a vortex avalanche cannot ever transfer so much angular momentum, that the crust rotates faster than the pinned superfluid; see \citet{Fulgenzi2017} and the output of Gross-Pitaevskii simulations \citep{Warszawski2011}}. This corresponds to a reflecting boundary condition at $X=0$:
\begin{equation}
\label{eq:refl}	
0 = \pd{p(X, t \,|\, X_0)}{X}\bigg\vert_{X=0} - \frac{2 \xi}{\sigma^2} p(X=0, t \,|\, X_0)\ \ . 
\end{equation}

\begin{figure}
\centering
\includegraphics[width=\linewidth]{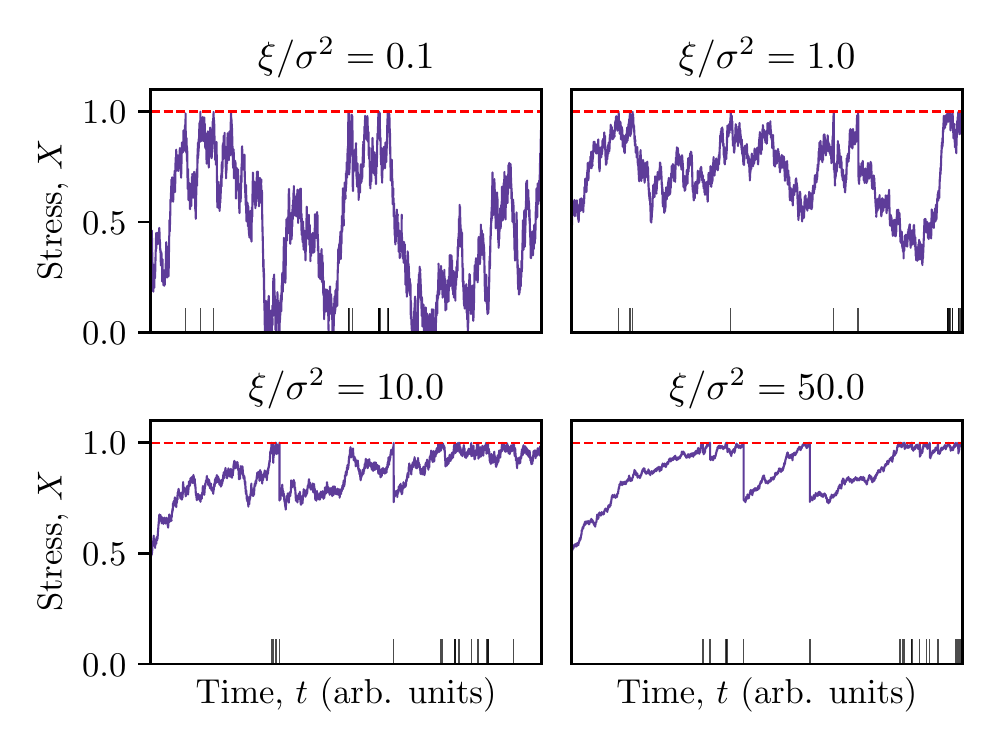}
\caption{Visual comparison of the evolution of the internal, unobservable stress, $X(t)$, for four different values of $\xi/\sigma^2$. The red dashed line indicates the stress threshold, set to $\xc=1$, where glitches are triggered. The same sequence of glitch sizes, drawn from a power-law $\eta(\Delta X)$ distribution, is used in each panel. The small black tick marks indicate the epoch of each glitch.}
\label{fig:mus_comp}
\end{figure}

Equations \eqref{eq:cke}--\eqref{eq:refl} are solved analytically assuming that $p(X, t \,|\, X_0)$ is separable in $X$ and $t$. The solution is presented in Appendix \ref{app:p_refl}, following the approach in \citet{Sweet1970}. Higher values of $\xi\xc/\sigma^2$ imply drift dominates over diffusion; lower values of $\xi\xc/\sigma^2$ imply diffusion dominates over drift. Figure \ref{fig:mus_comp} shows four representative time series of the evolution of $X$ for four different values of $\xi/\sigma^2$, with $\xc=1$ fixed in each panel. For $\xi / \sigma^2 = 0.1$ the process appears by eye to fluctuate randomly, with large, rapid excursions both up and down in stress. On the other hand, for $\xi / \sigma^2 = 50$, the stress accumulates steadily with small random excursions and large glitches are clearly demarcated from inter-glitch fluctuations.

\subsection{Waiting time and size distributions}
\label{sec:dist}
The stress is not observable. Instead, what we observe are sequences of glitch sizes and waiting times. 

The conditional waiting time distribution, $g(\Delta t \,|\, X_0)$, gives the probability density function (PDF) of waiting times $\Delta t$, when the inter-glitch evolution starts at $X_0$, according to \eqref{eq:ic}. It is calculated as \citep{Cox1965}
\begin{equation}
\label{eq:gderiv}
g(\Delta t \,|\, X_0) = -\frac{\td}{\td (\Delta t)} \left[\int_{-\infty}^{\xc}\td X\,  p(X, \Delta t \,|\, X_0) \right]\ \ .
\end{equation}
The integral inside the square brackets, often called the survivor function, equals the probability density that the process stays in the interval $-\infty < X(t) \leq \xc$ for $0 \leq t \leq \Delta t$.

The starting stress $X_0$ is a random variable, related to the size of the previous glitch. To find the observable waiting time distribution, $p(\Delta t)$, we marginalize over the starting stress by calculating,
\begin{equation}
\label{eq:pdelt}
p(\Delta t) = \int_0^{\xc} \td X_0\, g(\Delta t \,|\, X_0)\, \eta(\xc - X_0)\ \ ,
\end{equation}
where $\eta(\Delta X)$ equals the probability density of releasing an amount of stress $\Delta X = \xc - X_0$ during a glitch.

We henceforth express $t$ in units of $2\xc^2/\sigma^2$ and $X$ in units of $\xc$, unless otherwise stated. In these units, equations \eqref{eq:gderiv} and \eqref{eq:p_sol} combine to yield (see Appendix \ref{app:p_refl})
\begin{align}
\label{eq:grefl}
g(\Delta t \,|\, X_0) = &~ 2 \mu \exp\left[\mu^2 \Delta t + \mu (1 - X_0)\right] \nonumber\\
		&\times \sum_{n=1}^\infty \exp(-\lambda_n^2 \Delta t) \frac{\lambda_n \sin[\lambda_n(1 - X_0)]}{\mu + \cos^2\lambda_n}\ \ ,
\end{align}
where $\lambda_n$ is the $n$-th positive root of the transcendental equation
\begin{equation}
\label{eq:lamn}
 \mu \tan \lambda_n = -\lambda_n\ \ ,
\end{equation}
with
\begin{equation}
\mu = \xi \xc / \sigma^2\ \ .
\end{equation}

In this paper, we assume for simplicity that $\Delta X$ is proportional to the observed glitch size, $\Delta \nu$, i.e. the observed increment in the crust's spin frequency. Glitches represent small perturbations to an underlying equilibrium state, with $\Delta \nu / \nu \ll 1$, where $\nu$ is the spin frequency, so it is reasonable to model them in terms of a linear response, although nonlinear alternatives are certainly conceivable \citep{Alpar2006, Akbal2017}. In the vortex avalanche picture, for example, where $X(t)$ equals the crust-core angular velocity lag we have \citep{Fulgenzi2017}
\begin{equation}
\label{eq:delx}
\Delta X = -\frac{2\pi(I_\textrm{c} + I_\textrm{s}) \Delta \nu}{I_\textrm{s}}\ \ ,
\end{equation}
where $I_\textrm{c}$ and $I_\textrm{s}$ are the moments of inertia of the crust and superfluid interior respectively. An analogous proportionality exists in the starquake picture \citep{Middleditch2006, Chugunov2010}. The size distributions observed from individual pulsars are approximated by power-law, Gaussian, lognormal, and exponential distributions \citep{Melatos2008, Howitt2018, Fuentes2019}. Assuming $\Delta X \propto \Delta \nu$, we adjust $\eta(\Delta X)$ to match the measured size PDF $p(\Delta \nu)$ of the pulsar under consideration. 

\subsection{Average waiting time}
\label{sec:avet}
The average waiting time, $\avet$, is conditional on $X_0$. It can be calculated from $g(\Delta t \,|\, X_0)$ via
\begin{align}
\label{eq:avet}
\avet &= \int_0^\infty\td (\Delta t)\, \Delta t\, g(\Delta t \,|\, X_0) \ \ .
\end{align}
With the boundary conditions \eqref{eq:abs} and \eqref{eq:refl}, we obtain (see Appendix \ref{app:p_refl})
\begin{align}
\avet =&~ 2 \mu \exp\left[\mu(1 - X_0)\right] \nonumber\\
&\times\sum_{n=1}^\infty \left(\lambda_n^2 + \mu^2\right)^{-2}  \frac{\lambda_n \sin[\lambda_n(1 - X_0)]}{\mu + \cos^2(\lambda_n)}\ \ .
\end{align}
The behavior of $\avet$ as a function of $\mu$ is complicated, even after marginalizing over $X_0$. Numerical tests indicate that for $\mu \lesssim 1$, $\avet$ is roughly constant with $\mu$, while for $\mu \gtrsim 1$ it varies inversely with $\mu$. The latter behavior can be understood with the help of the approximate non-reflecting solution at large $\mu$ (see Appendix \ref{sec:norefl}), which has $\avet \propto \mu^{-1}$, via equation \eqref{eq:ig_delt} and \eqref{eq:avet}. The behavior at low values of $\mu$ makes sense physically, as $\sigma$ dominates the time to reach $\xc$ in this regime. On the other hand, at high values of $\mu$ and fixed $\sigma$, a high value of the drift coefficient $\xi$ leads the process to quickly reach $\xc$ while a low value of $\xi$ takes comparatively longer.

\subsection{Comparison with the SDP meta-model}
\label{sec:comp}
A key goal of this paper is to create a framework for falsifying one or both of the Brownian and SDP meta-models by making quantitative predictions about long-term glitch statistics. As the two meta-models encompass a range of plausible microphysics, falsifying one or both has significant scientific value in understanding which microphysical theories are consistent with the data.

The Brownian meta-model shares several similarities with the SDP meta-model \citep{Fulgenzi2017,Melatos2018,Carlin2019,Carlin2019b}. Both link the observed changes in $\nu(t)$ to a globally averaged, unobservable stress, which fluctuates around marginal stability. Both are examples of a self-organized critical system $[$see \citet{Aschwanden2018} for a review$]$, where an external driver pushes the system towards criticality, until a glitch releases internal stress and transfers angular momentum from the core to the crust \citep{Jensen1998}. Neither meta-model assumes a specific microphysical trigger mechanism; together the two meta-models embrace a wide variety of plausible mechanisms of stress accumulation and threshold triggering.

The meta-models also differ in important respects. The driver in the SDP meta-model is secular; it does not vary with time. In the Brownian meta-model the driving torque is a fluctuating Langevin torque with white noise statistics, as in \eqref{eq:sde}. The SDP process never quite reaches $X = \xc$, as glitches become increasingly likely for $X \rightarrow \xc$. In contrast, the Brownian meta-model reaches $X=\xc$ at every glitch. This has important implications regarding the ``memory'' of previous events, as explored in Section \ref{sec:correl}. Finally, $\eta(\Delta X)$ plays a different role in the two meta-models. As mentioned in Section \ref{sec:dist}, one has $\Delta \nu \propto \Delta X$, so $\eta(\Delta X)$ and $p(\Delta \nu)$ have the same shape in the Brownian meta-model. In the SDP meta-model $\eta(\Delta X)$ is conditional on $X(t)$ just before the glitch, so $\eta(\Delta X)$ and $p(\Delta \nu)$ have the same shape only under certain conditions; see \citet{Carlin2019} for details.

\begin{figure}
\centering
\includegraphics[width=\linewidth]{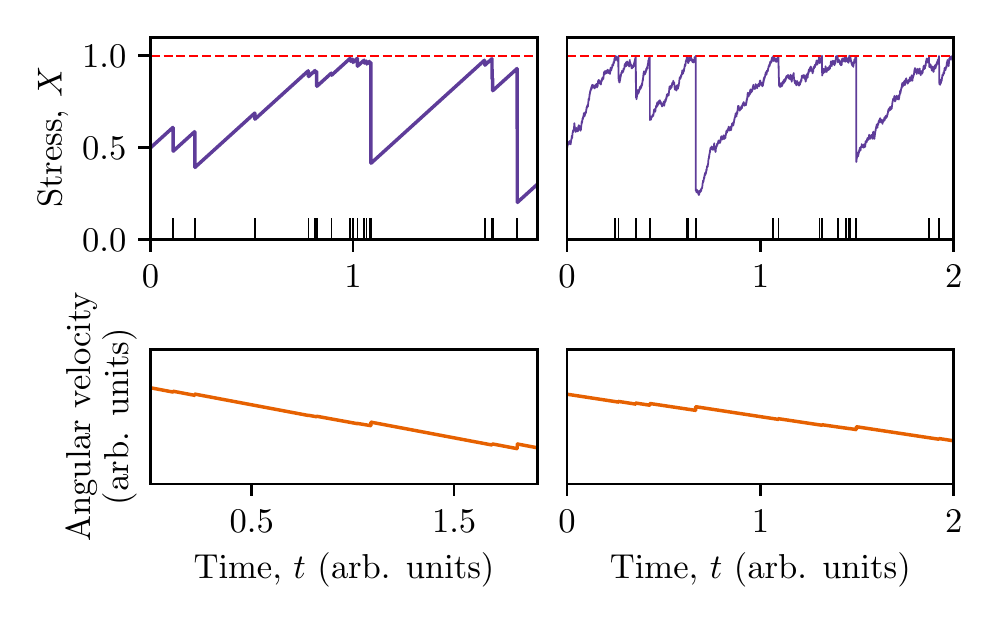}
\caption{Comparison between two representative time series of stress (top panels) and crust angular velocity (bottom panels) from the SDP meta-model (left) and the Brownian meta-model (right). A deterministic, secular torque drives the stress between glitches in the SDP meta-model, whereas a stochastic Langevin torque drives the stress between glitches in the Brownian meta-model. Black tick marks in the top panels indicate the glitch epochs. Parameters for SDP meta-model: $\alpha = 1$, power law conditional jump distribution, as described in equations (17) and (19) of \citet{Fulgenzi2017} respectively. Parameters for Brownian meta-model: $\mu=50$, power law stress-release distribution, as in \eqref{eq:eta_pow}. Parameters shared between meta-models: $\delta = -1.5$, $\beta=10^{-2}$ in \eqref{eq:eta_pow}.}
\label{fig:sd_comp}
\end{figure}

The similarities and differences between the two meta-models are illustrated in Figure \ref{fig:sd_comp}. Time series $X(t)$ and $\nu(t)$ are constructed by repeatedly evolving the stress in the system until a glitch is triggered (probabilistically at $X < \xc$ for the SDP meta-model, deterministically at $X=\xc$ for the Brownian meta-model), then drawing a glitch size from the stress-release PDF $\eta(\Delta X)$. Visually, with 20 glitches, the crust angular velocity evolves similarly for the two meta-models, despite the different stress evolution between glitches (deterministic for the SDP meta-model and stochastic for the Brownian meta-model). However, as we find in Section \ref{sec:results}, the long-term statistical behavior of the two meta-models is different.

\subsection{Inter-glitch spin wandering}
\label{sec:spinwand}
Besides its influence on glitch statistics, the Brownian process may also drive stochastic spin wandering between glitches, unlike the SDP process. In principle, therefore, observations of inter-glitch timing noise in radio pulsars \citep{Cordes1980, Price2012, Parthasarathy2019, Lower2020} should place constraints on the meta-model parameters $\xi$ and $\sigma^2$ independent of the constraints derived from glitches. As an illustrative special case, if $\xi$ and $\xc$ are held fixed, $\avet$ decreases and the inter-glitch timing noise amplitude increases simultaneously, as $\sigma^2$ increases. Hence a measured upper limit on the timing noise amplitude implies a maximum value of $\sigma^2$ and hence a minimum value of $\avet$, which provides an additional, independent opportunity to falsify the Brownian meta-model.

In practice, falsification experiments of the above kind are complicated by the unknown coupling between various components of the stellar interior. The meta-model parameters $\xi$ and $\sigma^2$ control the statistical behavior of the internal, i.e. unobservable, stress, $X(t)$. In Sections \ref{sec:eom} and \ref{sec:dist} we assume that changes in $X(t)$ couple linearly to the rotational frequency of the crust, $\nu(t)$, only when a glitch occurs, via \eqref{eq:delx}. If we relax this restriction and couple $X(t)$ linearly to the crust between glitches, we have
\begin{align}
\label{eq:ext_couple}
\frac{\td \nu}{\td t} = -A \frac{\td X}{\td t}\ ,
\end{align}
where $A$ is an unknown coupling constant (units: Hz per unit stress) which depends on the physical mechanism of stress accumulation and the microphysics controlling how the star's internal angular momentum reservoir is tapped in between glitches. Equation \eqref{eq:ext_couple} implies that, if the crust undergoes the same type of Brownian process with drift as described by \eqref{eq:sde}, the observable, long-term, average spin-down rate, $\langle\dot{\nu}\rangle$, is proportional to $\xi$, while the observed spin-wandering amplitude is proportional to $\sigma^2$. 

In the special case of $A=A_\textrm{max}$ (its maximum allowed value) the coupling is the same as during a glitch, e.g. $A = I_\textrm{s} / [2\pi (I_\textrm{c} + I_\textrm{s})]$ in the vortex avalanche picture. This is a problem for the Brownian meta-model, as we see from Figure \ref{fig:mus_comp}. To distinguish glitches from stochastic wandering we need $\mu \gtrsim 50$, otherwise large Brownian fluctuations can be mistaken for glitches. For $\mu \gtrsim 50$, there should be a strong cross-correlation between glitch sizes and waiting times until the next glitch, as discussed in Section \ref{sec:correl}. We do not see this cross-correlation in most pulsars, so we can rule out the special case of $A=A_\textrm{max}$ or the Brownian meta-model (or both).

On the other hand, for $A < A_\textrm{max}$, where the inter-glitch coupling is weaker than during a glitch, the problem outlined above is alleviated. Another scenario is that $A$ is not constant, i.e. it varies with time or the stress in the system. These scenarios are motivated by the observations of the ``precursor'' slowdown in the Vela pulsar immediately prior to the 2016 glitch \citep{Ashton2019}, and by studies of non-linear coupling mechanisms \citep{Akbal2017, Celora2020}. A detailed study of the microphysical implications of inter-glitch spin wandering for the coupling mechanism between the stress reservoir and the crust is left for future work. For simplicity, we assume henceforth that coupling only occurs at a glitch, via \eqref{eq:delx}.

\section{Observable long-term statistics}
\label{sec:results}
To prepare for comparing the Brownian meta-model to data, we study how changing the input parameters affects the long-term statistical predictions.  

\subsection{Waiting time distribution}
The long-term waiting time PDF, $p(\Delta t)$, constructed after many glitches are observed, is calculated from \eqref{eq:pdelt} given $\mu$ and $\eta(\Delta X)$. 
\begin{figure}
\centering
\includegraphics[width=\linewidth]{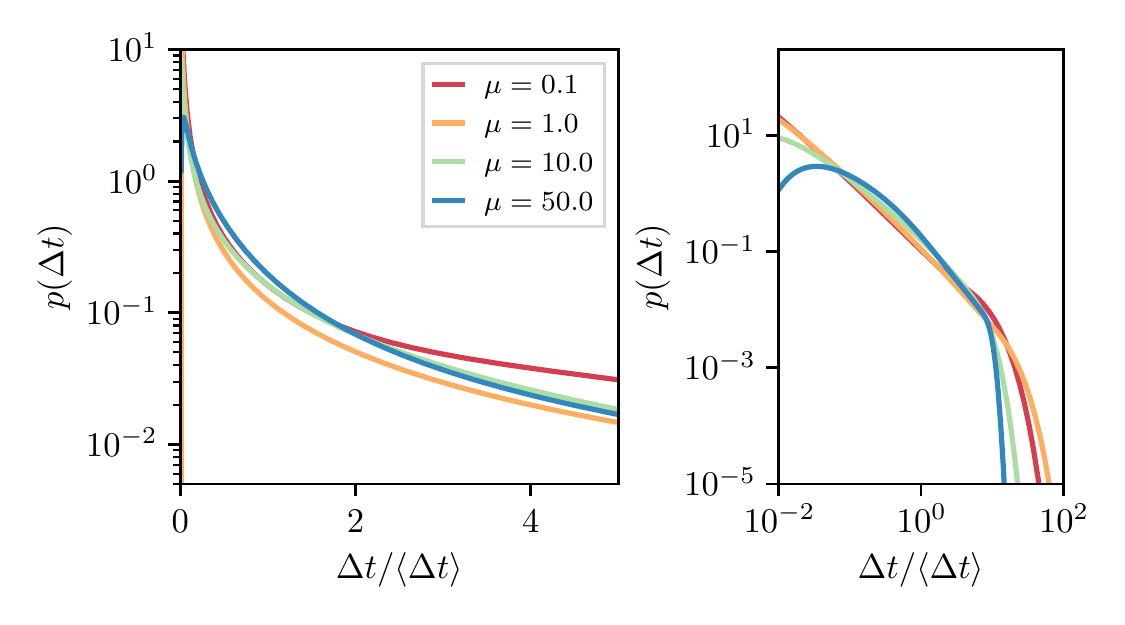}
\caption{Waiting time PDF, $p(\Delta t)$, for four values of $\mu$ on log-linear (left panel) and log-log (right panel) scales. The stress release distribution, $\eta(\Delta X)$, is a power law, as in \eqref{eq:eta_pow}, with $\delta=-1.5$ and $\beta=10^{-2}$.}
\label{fig:pdelt_pow}
\end{figure}
Figure \ref{fig:pdelt_pow} shows $p(\Delta t)$ for four representative values of $\mu$ when $\eta(\Delta X)$ is a power law of the form
\begin{equation}
\label{eq:eta_pow}
\eta(\Delta X) \propto \Delta X^{-\delta} H(1 - \Delta X) H(\Delta X - \beta)\ \ ,
\end{equation}
where the proportionality constant is fixed by $1 = \int_0^1 \td (\Delta X)\, \eta(\Delta X)$, $\delta$ is the power-law index, $\beta$ is the lower cut-off to ensure normalisability, and $H$ is the Heaviside function ($\beta \leq \Delta X \leq 1$ implies $0 \leq X \leq 1$ at all times). The abscissae are normalized by $\avet$ to highlight how the shape of $p(\Delta t)$ evolves with $\mu$. On the log-log axes (right panel) $p(\Delta t)$ resembles a power law over at least 3 decades, with a cut-off at $\Delta t \approx 10 \avet$. The cut-off steepens as $\mu$ grows. The shape of $p(\Delta t)$ depends weakly on $\delta$ and $\beta$ for $\mu \lesssim 10$, but depends strongly for $\mu \gtrsim 10$. For example, for $\mu \gtrsim 10$ and $\beta = 10^{-1}$, $p(\Delta t)$ becomes unimodal, as small waiting times become less likely when each glitch reduces the stress by $\Delta X \geq \beta$. 

\begin{figure}
\centering
\includegraphics[width=\linewidth]{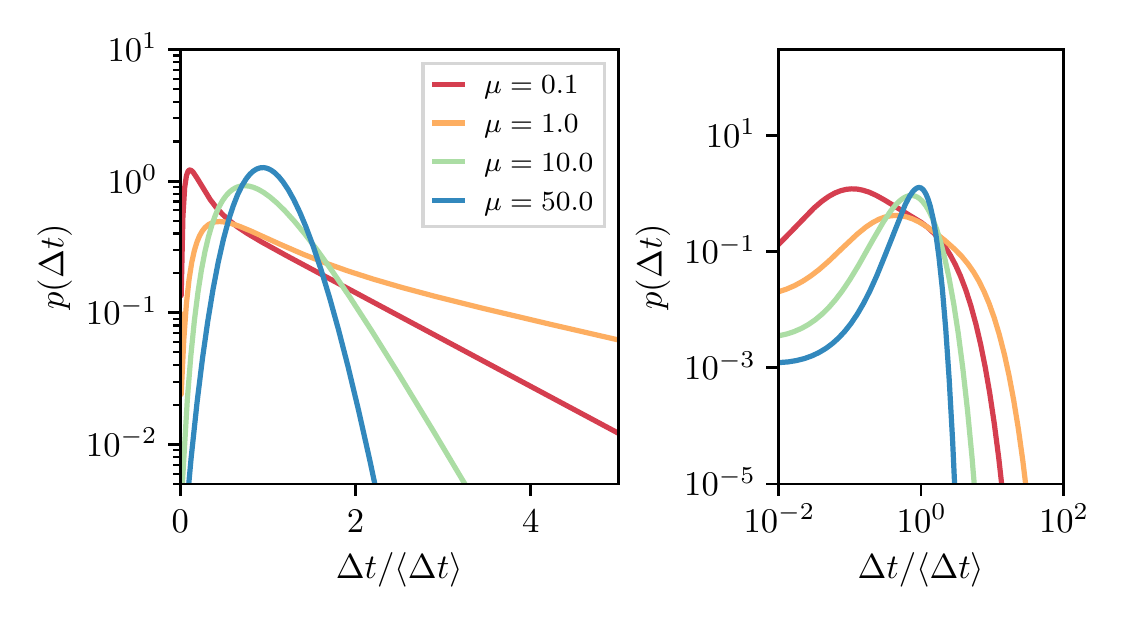}
\caption{As for Figure \ref{fig:pdelt_pow} but with a Gaussian $\eta(\Delta X)$, as in \eqref{eq:eta_gauss}, with $\mu_\textrm{G}=0.5$ and $\sigma_\textrm{G}=0.125$.}
\label{fig:pdelt_gauss}
\end{figure}

What about other functional forms of $\eta(\Delta X)$? Figure \ref{fig:pdelt_gauss} shows $p(\Delta t)$ for four representative values of $\mu$, with a Gaussian $\eta(\Delta X)$, viz.
\begin{equation}
\label{eq:eta_gauss}
\eta(\Delta X) \propto \exp\left[\frac{-\left(\Delta X - \mu_\textrm{G}\right)^2}{2\sigma_\textrm{G}^2}\right] H(1 - \Delta X) H(\Delta X)\ \ ,
\end{equation}
where the proportionality constant is fixed to normalize $\eta(\Delta X)$, $\mu_\textrm{G}$ is the mean, and $\sigma_\textrm{G}$ is the standard deviation. For $\mu \lesssim 1$, $p(\Delta t)$ resembles an exponential distribution, if the smallest waiting times with $\Delta t \lesssim 0.25 \avet$ are ignored. For $\mu \gtrsim 1$, $p(\Delta t)$ is unimodal. Increasing the size of the average $\Delta X$, via increasing $\mu_\textrm{G}$, reduces the variance in $p(\Delta t)$ for all $\mu$, whereas reducing $\mu_\textrm{G}$ makes $p(\Delta t)$ resemble the results for a power law $\eta(\Delta X)$. Reducing the variance of each stress-release event by reducing $\sigma_\textrm{G}$ also reduces the variance of $p(\Delta t)$, as expected.

\begin{figure}
\centering
\includegraphics[width=\linewidth]{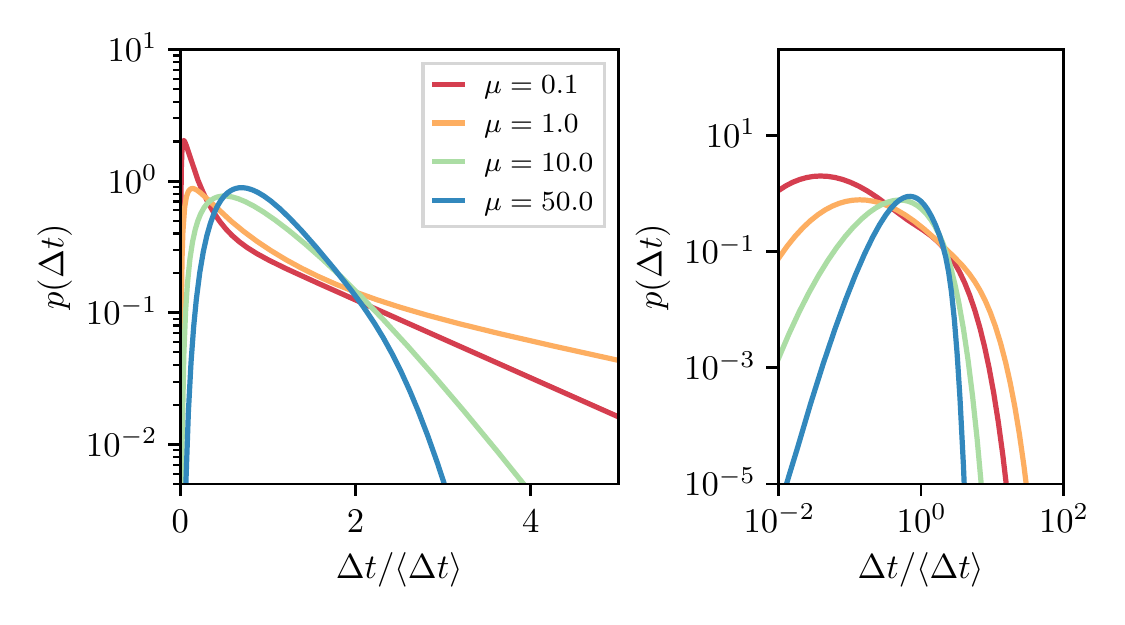}
\caption{As for Figure \ref{fig:pdelt_pow} but with a log-normal $\eta(\Delta X)$, as in \eqref{eq:eta_ln}, with $\mu_\textrm{LN}=-1$ and $\sigma_\textrm{LN}=0.5$.}
\label{fig:pdelt_ln}
\end{figure}

The third functional form of $\eta(\Delta X)$ that we test is a log-normal distribution,
\begin{equation}
\label{eq:eta_ln}
\eta(\Delta X) \propto \frac{1}{\Delta X} \exp\left[\frac{-(\log \Delta X - \mu_{\textrm{LN}})^2}{2\sigma^2_\textrm{LN}} \right] H(1 - \Delta X) H(\Delta X)\ \ ,
\end{equation}
where $\mu_\textrm{LN}$ and $\sigma_\textrm{LN}$ are the mean and standard deviation, and the proportionality constant is set by normalization. Figure \ref{fig:pdelt_ln} shows that the general shape of $p(\Delta t)$ with a log-normal $\eta(\Delta X)$ is similar to what is seen with a Gaussian $\eta(\Delta X)$. There are fewer small waiting times for a given $\mu$. If the average stress release is increased, by increasing $\mu_\textrm{LN}$, the same response is seen as with a Gaussian $\eta(\Delta X)$, i.e. the variance of $p(\Delta t)$ drops. If we increase $\sigma_\textrm{LN}$, $p(\Delta t)$ resembles what is seen with a uniform $\eta(\Delta X)$.

An analogous study of $p(\Delta t)$ for the SDP meta-model, with $\eta(\Delta X)$ taken to be a power law, Gaussian, and a variety of other functional forms, is presented by \citet{Carlin2019}.

\subsection{Correlations and memory}
\label{sec:correl}
The meta-model in Section \ref{sec:eom} predicts whether we should see a correlation between the size of a glitch and the subsequent waiting time, which we call a forward cross-correlation. As the glitch size is independent of the history of the stress evolution, there is no backward cross-correlation between the size of a glitch and the previous waiting time in the Brownian meta-model. Forward and backward cross-correlations have been investigated previously in the context of the SDP meta-model, and numerous falsifiable predictions are made \citep{Melatos2018, Carlin2019, Carlin2019b}. 

\begin{figure}
\centering
\includegraphics[width=0.8\linewidth]{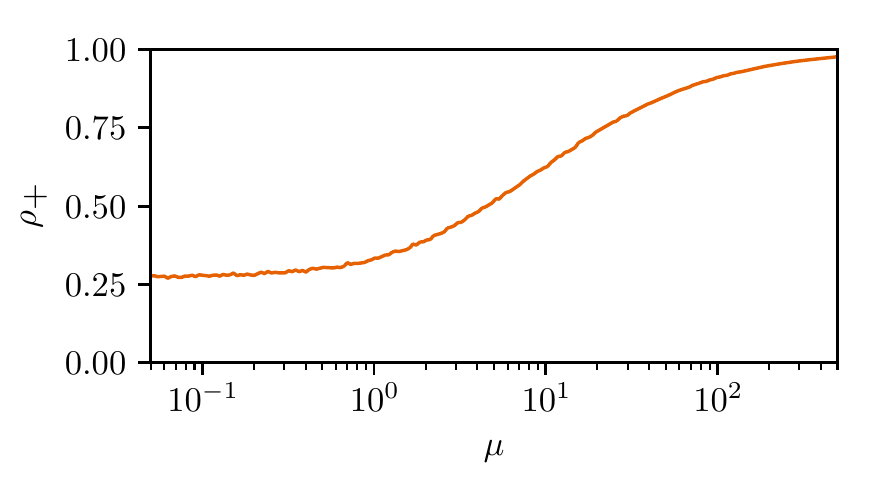}
\caption{Spearman cross-correlation between the size of a glitch and the waiting time until the next glitch versus $\mu$. At 200 logarithmically spaced values of $\mu$, $10^5$ waiting times and sizes are drawn from \eqref{eq:grefl} and \eqref{eq:eta_pow} respectively, using each generated size to determine the starting point for the next inter-glitch interval and hence waiting time.}
\label{fig:correl}
\end{figure}

Figure \ref{fig:correl} shows the Spearman correlation coefficient for the forward cross-correlation, $\rho_+$, for $5\times10^{-2} \leq \mu \leq 5\times10^3$. The cross-correlation is always positive and increases from $\rho_+ \approx 0.25$ for $\mu \lesssim 1$ to $\rho_+ \approx 1$ for $\mu \gg 1$. Figure \ref{fig:correl} is generated with $\eta(\Delta X)$ as a power law, but the result is insensitive to the form of $\eta(\Delta X)$. The trend in Figure \ref{fig:correl} is intuitive. The size of the stress release in a glitch dictates how much stress must be accumulated before the next glitch occurs. For $\mu$ high, the diffusion of the Brownian process is negligible compared to the secular drift, and so the waiting time is determined almost completely by the size of the previous glitch. For $\mu$ low, the diffusion randomizes the waiting time and decouples it from the size, while still maintaining a slight forward cross-correlation; even a process with zero drift is more likely to reach the threshold faster, if $X_0$ is closer to $\xc$. 
 
The Brownian meta-model predicts zero autocorrelations between glitch sizes, or between waiting times. The threshold at $X=1$ is reached before every glitch in the Brownian meta-model, removing ``memory'' in the system of the behavior of the stress prior to reaching that threshold. In contrast, the SDP meta-model predicts sizable autocorrelations in certain regimes \citep{Carlin2019b}.
 
\section{Falsifying the Brownian meta-model}
\label{sec:comp}
There are six pulsars with more than 15 recorded glitches\footnote{Up-to-date online catalogues of pulsar glitches are available through the Jodrell Bank Centre of Astrophysics at \url{http://www.jb.man.ac.uk/pulsar/glitches.html} \citep{Espinoza2011}, and the Australian National Telescope Facility at \url{https://www.atnf.csiro.au/research/pulsar/psrcat} \citep{Manchester2005}}. Their names, the number of recorded glitches, the forward Spearman cross-correlation coefficient (along with associated p-value and 95\% confidence interval), as well as the best-fitting size and waiting time distributions are listed in Table \ref{tab:data}. The Spearman correlation coefficient minimizes the impact of outliers by testing for monotonic correlations, as opposed to the strictly linear correlations which the standard Pearson correlation coefficient describes. The confidence interval is calculated as described in Section 4 of \citet{Carlin2019b}. The best-fitting PDFs are copied from \citet{Fuentes2019} and are selected based on the Akaike Information Criterion \citep{Akaike1974}. These shapes are broadly consistent with previous analyses using different techniques, although there are minor individual differences \citep{Melatos2008, Howitt2018}. We note that PDF shape fitting is uncertain when the sample size is small. Often the best one can do in the glitch context is to distinguish between a monotonic (e.g. exponential, power law) and unimodal (e.g. Gaussian) PDF, without tying down the functional form. Even then some functional forms (e.g. Weibull) straddle both categories \citep{Antonopoulou2018}. Further shape-fitting studies should be carried out in the future, as the data sets grow. 

Although not listed in Table \ref{tab:data}, we note that the backward cross-correlation, the autocorrelation between glitch sizes, and the autocorrelation between waiting times are all consistent with zero, at a 95\% confidence level for all six objects \citep{Melatos2018, Carlin2019b, Fuentes2019}.

\begin{table*}
	\caption{Pulsar name, number of glitches ($N$), Spearman correlation coefficient between glitch size and subsequent waiting time ($\rho_+$), associated p-value, and the 95\% confidence interval (CI). The last two columns indicate the functional form of the best-fitting distribution for glitch sizes, $p(\Delta X)$, and waiting times, $p(\Delta t)$ \citep{Melatos2008, Howitt2018, Fuentes2019}.}
	\begin{center}
		\begin{tabular}{l c l l l l l}
		\toprule
		Name (PSR J) 	& $N$ &  {$\rho_+$}	& p-value & 95\% CI & $p(\Delta X)$ & $p(\Delta t)$\\
		\midrule
		0537$-$6910  & 42$^*$ & 0.93  & $10^{-18}$  & (0.84, 0.97) & Gaussian & Gaussian \\
		1740$-$3015  & 36 & 0.29  & 0.091 & ($-$0.06, 0.58) & Power law & Exponential \\
		0534$+$2200  & 25$^\dagger$ & $-$0.060 & 0.78  & ($-$0.45, 0.35) & Log-normal & Exponential \\
		1341$-$6220  & 23 & 0.58  & 0.0048  & (0.13, 0.83) & Log-normal & Exponential \\
		0835$-$4510  & 21 & 0.30  & 0.20  & ($-$0.19, 0.67) & Gaussian & Gaussian \\
		0631$+$1036  & 17 & 0.21  & 0.44  & ($-$0.33, 0.65) & Power law & Exponential\\
		\bottomrule
		\end{tabular}
	\end{center}
	\label{tab:data}
	\justify
	$^*$The number and parameters of glitches in PSR J0537$-$6910 vary between \citet{Middleditch2006}, \citet{Antonopoulou2018}, and \citet{Ferdman2018}. We include in our analysis glitches that appear in two out of three sources.
	\newline$^\dagger$The first four PSR J0534$+$2200 glitches in the Jodrell Bank catalogue occurred before daily monitoring commenced and are excluded from the analysis \citep{Lyne2015}.
\end{table*}	

One virtue of the Brownian meta-model, like the SDP meta-model studied elsewhere \citep{Fulgenzi2017, Melatos2018, Carlin2019, Carlin2019b}, is that it makes specific, quantitative predictions about PDFs and correlations. These predictions are open to falsification using existing and future data. With an eye to falsifying the meta-model presented in Sections \ref{sec:bm_meta} and \ref{sec:results} we now ask whether existing long-term observations of the pulsars in Table \ref{tab:data} can be adequately explained. In doing so, we caution that there is debate about whether the existing glitch catalogues are complete and accurate. \citet{Espinoza2014} claimed that all glitches in the Crab pulsar (PSR J0534$+$2200) are detected. \citet{Yu2017b} used a Monte Carlo study to confirm that the \citet{Yu2013} analysis of 165 pulsars observed between 1990 and 2011 using the Parkes Observatory has ``detected all detectable glitches in the data'' (verbatim quote). However, as the cadence of observations for most pulsars is not constant \citep{Janssen2006}, post-glitch recovery time-scales vary \citep{Alpar1984,VanEysden2010}, and glitch detections still rely on human intervention \citep{Yu2017b}, it remains uncertain whether or not we are seeing the smallest glitches, or resolving glitches that happen in quick succession.

\subsection{PSR J0537$-$6910}
PSR J0537$-$6910 has the most recorded glitches and the highest forward cross-correlation amongst all the prolific glitching pulsars. In the context of the Brownian meta-model, these properties place PSR J0537$-$6910 in the $\mu \gtrsim 10^2$ regime, via Figure \ref{fig:correl}. The glitch size distribution for PSR J0537$-$6910 is approximately Gaussian \citep{Fuentes2019}. Looking at Figure \ref{fig:pdelt_gauss}, where $\eta(\Delta X)$ is a Gaussian, we note that $p(\Delta t)$ should also be a Gaussian, with $\mu \gtrsim 10^2$ , in accord with observations. Therefore, the main features of the long-term statistics of this pulsar conform to the Brownian meta-model, if $\eta(\Delta X)$ is a Gaussian, and one has $\mu \gtrsim 10^2$.

We note that the waiting time distribution for PSR J0537$-$6910 is also well described by a Weibull distribution \citep{Antonopoulou2018}, a more general functional form, which includes the exponential and a skewed Gaussian as special cases. 

\subsection{PSR J1740$-$3015}
PSR J1740$-$3015 has a forward cross-correlation that is consistent with zero. However the 95\% confidence interval is broad enough to encompass $\rho_+$ up to 0.58. According to Figure \ref{fig:correl} this means PSR J1740$-$3015 has $\mu \lesssim 10$, in the context of the Brownian meta-model. As PSR J1740$-$3015 has a power-law size PDF \citep{Fuentes2019}, we look to Figure \ref{fig:pdelt_pow}, where $\eta(\Delta X)$ is a power law. For $\mu \lesssim 10$ the Brownian meta-model predicts that $p(\Delta t)$ is a power law with a cut-off at large $\Delta t$. Therefore, as $p(\Delta t)$ is observed to be exponential in this object, the long-term statistics are not explained by the Brownian meta-model with any set of input parameters.

Power-law and log-normal distributions are often hard to distinguish for such small sample sizes. If $\eta(\Delta X)$ is actually a log-normal distribution for this object, then we look at Figure \ref{fig:pdelt_ln}. With $\mu \lesssim 10$ we note that $p(\Delta t)$ should be an exponential, if the smallest waiting times are not observed. Therefore, as $p(\Delta t)$ is observed to be exponential in this object, it is consistent with the Brownian meta-model, if we are unable to observe glitches with $\Delta t \lesssim 0.25 \avet$. Note that $\avet$ refers to the true underlying average waiting time, rather than the estimate from the sample of glitches we have observed. 

\subsection{PSR J0534$+$2200}
PSR J0534$+$2200 has a forward cross-correlation that is consistent with zero, with $\rho_+ \leq 0.35$ at 95\% confidence. This limits PSR J0534$+$2200 to $\mu \lesssim 2$, according to Figure \ref{fig:correl}. PSR J0534$+$2200 has a log-normal size distribution \citep{Fuentes2019}. Taking $\eta(\Delta X)$ to be log-normal, as in Figure \ref{fig:pdelt_ln}, we see that $p(\Delta t)$ should be an exponential, if the smallest waiting times are not observed. Therefore, as $p(\Delta t)$ is observed to be exponential in this object, it is consistent with the Brownian meta-model, if we are unable to observe glitches with $\Delta t \lesssim 0.25 \avet$. If we do see all glitches in PSR J0534$+$2200, as claimed by \citet{Espinoza2014}, then the observations are inconsistent with the Brownian meta-model.

We note that the semi-autonomous glitch-finding algorithm of \citet{Espinoza2014} may miss closely spaced glitches occasionally. For example, it missed one glitch, at epoch MJD 52146.8 with a size of $\Delta \nu=0.27\,\mu$Hz, which occurred $\Delta t \approx 63\,$d after the previous glitch with \citet{Espinoza2014} noting that the likely cause is ``influence of the recovery from the previous glitch'' (verbatim quote). If we take 63$\,$d as the minimum resolvable waiting time, the true underlying average waiting time is $\avet \approx 63\,\textrm{d} / 0.25 = 252\,$d, in order for the long-term statistics to be consistent with the Brownian meta-model. The observed average waiting time is 501$\,$d, while the median waiting time is 284$\,$d. On the other hand, the Brownian meta-model may be ruled out, and the minimum resolvable waiting time may be shorter than $63\,$d. More work is needed to clarify these issues, including systematic studies of the false alarm and false dismissal probabilities of glitch-finding algorithms \citep{Janssen2006, Shannon2016, Yu2017b, Lower2020}.

\subsection{PSR J1341$-$6220}
PSR J1341$-$6220 has a forward cross-correlation that is significantly positive. However the 95\% confidence interval is broad, allowing $0.13 \leq \rho_+ \leq 0.83$. According to Figure \ref{fig:correl} this limits $\mu$ to $\mu \lesssim 10^2$. PSR J1341$-$6220 has a log-normal size distribution \citep{Fuentes2019}, and so like PSR J0534$+$2200 is consistent with Brownian the meta-model, only if we do not detect glitches with $\Delta t \lesssim 0.25\avet$. The observed waiting time distribution is an exponential.

\subsection{PSR J0835$-$4510}
PSR J0835$-$4510 has a forward cross-correlation that is consistent with zero. The 95\% confidence interval encompasses $\rho_+$ up to 0.67, consistent with $\mu \lesssim 30$, according to Figure \ref{fig:correl}. The size PDF, and hence $\eta(\Delta X)$, for PSR J0835$-$4510 is approximately Gaussian \citep{Fuentes2019}. Therefore according to Figure \ref{fig:pdelt_gauss} the meta-model predicts $p(\Delta t)$ to be an exponential (for $\mu \lesssim 5$) or a skewed Gaussian (for $5 \lesssim \mu \lesssim 30$). The observed $p(\Delta t)$ is a Gaussian, not an exponential. Therefore, the observations are currently consistent with the Brownian meta-model for $5 \lesssim \mu \lesssim 30$, if $\eta(\Delta X)$ is a Gaussian. 

The somewhat strict constraints on $\mu$ imply that, with more glitches, the measured forward cross-correlation should increase to $0.4 \lesssim \rho_+ \lesssim 0.6$. If $\rho_+$ stays outside this range, PSR J0835$-$4510 will become another counterexample to the Brownian meta-model.

\subsection{PSR J0631$+$1036}
PSR J0631$+$1036 has roughly half the recorded glitches of PSR J1740$-$3015 but is otherwise similar statistically. Hence the same conclusion holds: as long as the size distribution is a power law \citep{Fuentes2019}, the Brownian meta-model does not adequately explain the observations, as exponential waiting times cannot be generated if $\eta(\Delta X)$ is a power law.

As with PSR J1740$-$3015, if $\eta(\Delta X)$ is actually a log-normal distribution, instead of a power law, the conclusion is different: the observations are consistent with the predictions of the Brownian meta-model, if we do not resolve glitches with $\Delta t \lesssim 0.25 \avet$.

\section{Population trends}
\label{sec:poptrends}
The primary goal of this paper is to formulate rigorously and then falsify (if possible) the Brownian meta-model, rather than engage in a parameter estimation exercise. Nevertheless the results in Section \ref{sec:comp} do carry some interesting preliminary implications concerning the parameters of the Brownian meta-model, in the event that it survives falsification in the future. In this section, we touch briefly on two population trends that are consistent with (albeit not guaranteed by) the results in Section \ref{sec:comp}: why do $\eta(\Delta X)$ and $\mu$ seem to vary significantly among the six pulsars in Table \ref{tab:data}?

Regarding $\eta(\Delta X)$, laboratory studies of self-organized critical systems with avalanche dynamics, like sand piles, reveal that $\eta(\Delta X)$ is power-law-like when the driver is ``slow'', and Gaussian-like when the driver is ``fast'' \citep{Jensen1998}. In the former regime, avalanches occur sporadically at well-separated points within the system, so consecutive avalanches are independent and scale invariant: they can have any size, ranging from a solitary nearest-neighbor interaction to a catastrophic collapse of the whole system. In the latter regime, consecutive avalanches ``trip over one another'' (i.e. are correlated, not independent) and involve most of the system every time, so they all have comparable sizes, and $\eta(\Delta X)$ is unimodal. Broadly speaking the foregoing physics may suggest a correlation between the shape of $\eta(\Delta X)$ and $\langle \dot{\nu} \rangle$, and it will be interesting to test for such a correlation in the future, as more data are gathered. However, one must approach such a test with caution. The demarcation between ``slow'' and ``fast'' drivers is a subtle and unsolved question in idealized systems like sand piles, let alone in neutron stars where the microphysics is complicated and unknown (e.g. vortex avalanches, starquakes). Moreover observables like $\langle \dot{\nu} \rangle$ cannot be related easily to the behavior of the stress reservoir, e.g. due to uncertain coupling between multiple components of the star's interior, as discussed in Section \ref{sec:spinwand}.

To understand how $\mu = \xi \xc / \sigma^2$ could vary pulsar-to-pulsar we need to unpack the various internal parameters, and relate them to potential observables. In the standard picture, $\langle \dot{\nu} \rangle$ is set by the spin-down torque, $N_\textrm{ext}$, and moment of inertia of the crust, $I_\textrm{c}$. As discussed in Section \ref{sec:spinwand}, one can invoke a linear coupling between the internal stress and observed behavior of the crust. Linear coupling faces many issues, as we discuss in Section \ref{sec:spinwand}, but taking it to be valid for the moment, we find $\xi \propto \langle\dot{\nu}\rangle \approx N_\textrm{ext} / I_\textrm{c}$, where the proportionality constant controls the strength of the coupling. For the six objects discussed in this paper, $N_\textrm{ext} \propto B^2 \nu^3$ (where $B$ is the strength of the dipole magnetic field at the surface) varies across three orders of magnitude, using values of $B$ and $\nu$ from the ATNF pulsar catalogue\footnote{\url{https://www.atnf.csiro.au/research/pulsar/psrcat/} \citep{Manchester2005}}. The other factor is $I_\textrm{c}$. There are two popular scenarios for this quantity, as discussed in Section 3 of \citet{Melatos2015}: (a) if the crust is a thin crystalline lattice and the rest of the star is composed of a superfluid we have $I_\textrm{c} / I_0 \sim 10^{-2}$, where $I_0$ is the total moment of inertia of the star \citep{Andersson2012, Hooker2015}; (b) if the crust has most of the interior superfluid pinned and co-rotating with it (via magnetic flux tubes or charged particles), with only a bit of the inner crust superfluid decoupled, we have $I_\textrm{c} / I_0 \sim 1$ \citep{Link1999a, Lyne2000a, Espinoza2011}. We do not explore which of these scenarios is more likely, as both have strong support in the literature. We do note that the difference between these scenarios widens the possible range of $\xi$ by another two orders of magnitude. The other factors in $\mu$ are $\sigma$ and $\xc$. Again, as discussed in Section \ref{sec:spinwand}, $\sigma$ is proportional to the observed spin-wandering amplitude, if we assume a linear coupling. The spin-wandering amplitude in the six objects considered in this paper is not well quantified in the literature. However, \citet{Shannon2010} found that for a general population of ``canonical pulsars'', the timing noise strength, $\sigma_{\textrm{TN}}$, spans three orders of magnitude. Finally, the critical stress $\xc$ may vary from object to object, as it is a complex combination of microphysical (e.g. pinning potential)  and thermodynamic (e.g. equation of state) parameters \citep{Link1991}. Hence, even for linear coupling (which is already ruled out by looking at inter-glitch spin wandering, as discussed in Section \ref{sec:spinwand}), the possible range of $\mu$ inferred from external observables spans more than eight orders of magnitude, comfortably encompassing the range of $\mu$ which the meta-model considers.

\section{Conclusions}
The physical mechanism that triggers pulsar glitches is unknown. Phenomenological meta-models offer one way to link --- and potentially falsify --- broad classes of plausible microphysical mechanisms with measurements of long-term glitch statistics. The SDP meta-model \citep{Fulgenzi2017} describes microphysical mechanisms in which glitches are triggered probabilistically, while the stress in the system rises secularly, becoming more likely as the stress increases. It makes falsifiable, quantitative predictions for size and waiting-time cross-correlations \citep{Melatos2018}, autocorrelations \citep{Carlin2019b}, and PDFs \citep{Carlin2019}. However, the SDP meta-model does not allow the stress to fluctuate stochastically in between glitches due to random processes in the stellar interior, e.g. superfluid vortex motion \citep{Warszawski2011}, superfluid turbulence \citep{Melatos2010, Melatos2014}, or crust cracking \citep{Horowitz2009}. 

Motivated partly by recent observations of PSR J0835$-$4510 \citep{Ashton2019}, we introduce an alternative meta-model, where the stress evolves between glitches according to a Brownian process with drift and diffusion components, and where glitches are triggered deterministically once the stress surmounts a threshold. The rotational slowdown observed by \citet{Ashton2019} just prior to the glitch may be a coincidently large instance of pulse jitter, but it may also indicate a large, stochastic fluctuation in the internal stress, which briefly couples the magnetosphere to the interior and triggers the glitch. While we do not model the microphysics in detail, the Brownian meta-model encompasses such a trigger mechanism. We show in Section \ref{sec:comp} and \citet{Carlin2019b} that the glitch statistics of PSR J0835$-$4510 are consistent with the predictions of both the Brownian and SDP meta-models.

We find that the Brownian meta-model predicts various long-term statistical fingerprints. If the glitch size distribution is not a power law, and diffusion dominates drift (i.e. $\mu \lesssim 1$), the waiting time PDF is predicted to be an exponential, if glitches that occur soon after one another are not resolved. As $\mu$ increases, the observed waiting time PDF resembles more closely the glitch size PDF. The Spearman cross-correlation coefficient between glitch size and waiting time until the next glitch is predicted to be at least 0.25 for all pulsars.

Current observations of the long-term glitch statistics in all six of the pulsars with the most recorded glitches cannot be explained adequately by the Brownian meta-model. The two ``quasi-periodic'' glitchers (PSR J0537$-$6910 and PSR J0835$-$4510) with Gaussian size and waiting time distributions \citep{Howitt2018, Fuentes2019} can be explained with the Brownian meta-model, while PSR J1740$-$3015 and PSR J0631$+$1036 cannot (regardless of input parameters), unless their glitch sizes are distributed as a log-normal instead of a power law \citep{Fuentes2019}. PSR J0534$+$2200 and PSR J1341$-$6220 are consistent with the meta-model, if there are many glitches with small waiting times that we do not resolve. More data could falsify the Brownian meta-model as it applies to individual pulsars in several ways: \begin{enumerate*} \item if the measured forward cross-correlation is statistically inconsistent with $\rho_+ \geq 0.25$; \item if a non-zero backward cross-correlation is measured; or \item if the size or waiting time autocorrelations are nonzero.\end{enumerate*} Additionally, measurements of the forward cross-correlation, combined with the size and waiting time PDFs, further constrain the meta-model parameters.

We note that \begin{enumerate*} \item the SDP meta-model is broadly consistent with the long-term statistics in the six pulsars with the most recorded glitches \citep{Carlin2019, Carlin2019b}, and \item it predicts a different set of long-term statistics.\end{enumerate*} Thus, over time we can distinguish between the two meta-models and falsify one, the other, or both. We remind the reader that most plausible microphysical mechanisms contemplated in the literature (e.g. superfluid vortex avalanches, starquakes, hydrodynamic instabilities and turbulence) fit broadly within one or both of the Brownian and SDP meta-models.

\section*{Acknowledgements}
Parts of this research are supported by the Australian Research Council (ARC) Centre
of Excellence for Gravitational Wave Discovery (OzGrav) (project number
CE170100004) and ARC Discovery Project DP170103625. JBC is supported by an Australian Postgraduate Award. We thank the anonymous referee for pointing out that inter-glitch spin wandering places independent constraints on the Brownian meta-model in principle, as discussed in Section \ref{sec:spinwand}.



\newpage
\bibliographystyle{mnras}
\bibliography{zot_bib_bm}



\appendix
\section{Analytic solution of the Fokker-Planck equation for the inter-glitch stress distribution}
\label{app:p_refl}

The Fokker-Planck equation for the globally averaged stress variable, $X$, together with the initial and boundary conditions, \eqref{eq:ic}--\eqref{eq:refl}, constitute a standard diffusion problem. Namely, equation \eqref{eq:cke} is a parabolic partial differential equation with constant coefficients solved on the finite interval $0 \leq X \leq \xc$, subject to mixed Dirichlet-Neumann (also called Robin) boundary conditions. The problem can be solved analytically by expanding the solution in eigenfunctions on the interval $0 \leq X \leq \xc$ \citep{Sweet1970}.

We assume a separable ansatz 
\begin{equation}
p(X, t) = Y(X) T(t)\ \ ,
\end{equation}
which converts \eqref{eq:cke} into two coupled ordinary differential equations,
\begin{align}
\label{eq:t_orig}
\frac{2}{\sigma^2 T} \frac{\td T}{\td t} &= -\alpha^2\ \ ,\\
\frac{1}{Y} \left(\frac{-2\xi}{\sigma^2} \frac{\td Y}{\td X} + \frac{\td^2 Y}{\td X^2} \right) &= -\alpha^2\ \ , \label{eq:y_ode}
\end{align}
for some constant $\alpha$. Equation \eqref{eq:y_ode} has exponential solutions of the form
\begin{equation}
Y(X) \propto \exp\left[\left(\frac{\xi}{\sigma^2} \pm \sqrt{-\lambda^2}\right) X\right]\ \ ,
\label{eq:ysol}
\end{equation}
with $\lambda^2 = \alpha^2 - \xi^2/\sigma^4$. 

As \eqref{eq:cke} is linear, we apply the boundary conditions to the eigenfunctions defined in \eqref{eq:ysol} independently, then sum over the eigenvalues using the principle of superposition. For $\lambda^2 \leq 0$, $Y(X)$ becomes a linear combination of $\sinh(\lambda X)$ and $\cosh(\lambda X)$. The boundary conditions imply $\tanh(\lambda X) \propto -\lambda$, whose only solution $\lambda = 0$ leads to the trivial result $Y(X) = 0$. We therefore restrict our attention to $\lambda^2 > 0$ and hence
\begin{equation}
Y(X) = \exp\left(\frac{\xi}{\sigma^2}X\right) \left(A \sin\lambda X + B \cos\lambda X \right)\ \ ,
\end{equation}
where $A$ and $B$ are constants. The reflecting boundary condition \eqref{eq:refl} implies
\begin{align}
B &= \frac{\lambda \sigma^2}{\xi} A\ \ ,
\end{align}
while the absorbing boundary condition \eqref{eq:abs} fixes the eigenvalues, $\lambda$, via
\begin{align}
\tan(\lambda \xc) &= -\frac{\lambda \sigma^2}{\xi}\ \ .\label{eq:lam_non}
\end{align}
Hence we write the full solution for $P(X,t)$ as
\begin{align}
P(X,t) = &~\exp\left(\frac{\xi}{\sigma^2}X\right) \sum_{n=1}^\infty A_n \exp \left[-t \left(\frac{\lambda_n^2 \sigma^2}{2} + \frac{\xi^2}{2\sigma^2} \right)\right] \nonumber\\ 
&~ \times \left[\sin(\lambda_n X) + \frac{\lambda_n \sigma^2}{\xi} \cos(\lambda_n X) \right]
\end{align}
or equivalently
\begin{align}
P(X,t) = &~\exp\left(\frac{\xi}{\sigma^2}X\right) \sum_{n=1}^\infty A_n' \exp \left[-t \left(\frac{\lambda_n^2 \sigma^2}{2} + \frac{\xi^2}{2\sigma^2} \right)\right] \nonumber\\ 
&~ \times \sin[\lambda_n (X - \xc)]\ \ , \label{eq:p_sol}
\end{align}
where $\lambda_n$ is the $n$-th positive root of \eqref{eq:lam_non}, and the $A_n'$ constant coefficients are to be determined.

We find the $A_n'$ factors by applying the initial condition \eqref{eq:ic} and noting that the eigenfunctions are orthogonal on $0 \leq X \leq \xc$ (not the standard Fourier domain $0 \leq X \leq 2\pi$) as a consequence of Sturm-Liouville theory \citep{Morse1953}. Orthogonality implies
\begin{align}
A_n' = &~\frac{\int_0^{\xc} \td X\, \exp\left(-\xi X / \sigma^2\right)\sin[\lambda_n(X - \xc)]\, p(X, t=0)}{\int_0^{\xc}\td X\, \sin^2[\lambda_n(X - \xc)]}\\
= &~2 \exp\left(-\xi X_0 / \sigma^2\right)\sin[\lambda_n(X_0 - \xc)]\nonumber\\ &~\times \frac{\xi/\sigma^2}{\xi \xc / \sigma^2 + \cos^2(\lambda_n \xc)}\ \ .
\label{eq:an_p}
\end{align}
The full solution is given by \eqref{eq:lam_non}, \eqref{eq:p_sol}, and \eqref{eq:an_p}.

\section{Conditional waiting time PDF without the reflecting boundary}
\label{sec:norefl}
If the reflecting boundary condition \eqref{eq:refl} is relaxed, such that the process operates on the semi-infinite domain $X < \xc$, the conditional waiting time distribution is an inverse Gaussian \citep{Cox1965}, 
\begin{equation}
\label{eq:ig_delt}
g(\Delta t \,|\, X_0) = \frac{\xc - X_0}{\sigma\sqrt{2\pi \Delta t^{3}}} \exp \left[ \frac{-(\xc - X_0 - \xi \Delta t)^2}{2 \sigma^2 \Delta t} \right]\ \ .
\end{equation}
For $\xi/\sigma^2 \gtrsim 10$, numerical tests show that \eqref{eq:ig_delt} agrees with \eqref{eq:grefl} to within 1\% for $0 \leq \Delta t \leq 5 \avet$. This makes intuitive sense, as the process is driven strongly away from $X = 0$ for large $\xi/\sigma^2 > 0$. We use \eqref{eq:ig_delt} instead of \eqref{eq:grefl} for $\xi/\sigma^2 \gtrsim 10$, because \eqref{eq:grefl} converges slowly in the latter regime.


\bsp	
\label{lastpage}
\end{document}